%% file: main.tex
\pdfoutput=1
\PassOptionsToPackage{table}{xcolor}

\documentclass[11pt]{article}

\usepackage[preprint]{acl}

\usepackage{times}
\usepackage{latexsym}
\usepackage{multirow}
\usepackage{booktabs}
\usepackage{pifont} % For checkmarks and crosses
\usepackage{xspace}
\usepackage{amsmath}
\usepackage{adjustbox} % For adjusting the size of the table
\newcommand{\cmark}{\ding{51}} % Checkmark
\newcommand{\xmark}{\ding{55}} % Cross
\usepackage{listings}

\definecolor{lightgray}{rgb}{0.95, 0.95, 0.95}
\definecolor{darkgray}{rgb}{0.4, 0.4, 0.4}
\definecolor{backcolour}{rgb}{0.95,0.95,0.92}
\definecolor{myblue}{rgb}{0.2, 0.4, 0.8} % Example blue color
\definecolor{mygreen}{rgb}{0.2, 0.6, 0.2} % Example green color
\usepackage{array}
\newcolumntype{L}[1]{>{\raggedright\arraybackslash}p{#1}}

\newcommand{\framework}[1]{\textsc{#1}\xspace}

\newcommand{\ourframework}{\framework{Fire}}
\newcommand{\safe}{\framework{Safe}}

\newcommand{\factool}{\framework{FacTool}}
\newcommand{\factcheckgpt}{\framework{Factcheck-GPT}}

\newcommand{\model}[1]{\text{#1}\xspace}
\newcommand{\fouro}{\model{GPT-4o}}
\newcommand{\fouromini}{\model{GPT-4o-mini}}
\newcommand{\oone}{\model{o1-preview}}
\newcommand{\oonemini}{\model{o1-mini}}
\newcommand{\llama}{\model{LLaMA}}

\newcommand{\haiku}{\model{Claude-3 Haiku}}
\newcommand{\opus}{\model{Claude-3 Opus}}
\newcommand{\sonnet}{\model{Claude-3.5 Sonnet}}
\newcommand{\mistral}{\model{Mistral}}

\newcommand{\dataset}[1]{\text{#1}\xspace}

\newcommand{\factoolqa}{\dataset{FacTool-QA}}
\newcommand{\felmwk}{\dataset{FELM-WK}}
\newcommand{\factcheckbench}{\dataset{Factcheck-Bench}}

\newcommand{\factbench}{\dataset{FactBench}}
\newcommand{\bingcheck}{\dataset{BingCheck}}

\definecolor{Mulberry}{rgb}{0.77,0.29,0.55}
\definecolor{CadmiumOrange}{rgb}{0.93,0.53, 0.18}
\definecolor{ForestGreen}{rgb}{0.13, 0.55, 0.13}
\definecolor{LimeGreen}{RGB}{50, 205, 50}

\usepackage[T1]{fontenc}
\usepackage[utf8]{inputenc}

\usepackage{microtype}

\usepackage{inconsolata}

\usepackage{graphicx}

\title{\ourframework\includegraphics[height=1.5em]{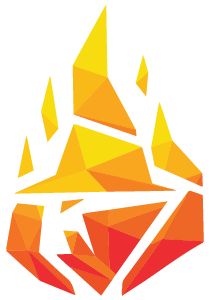}: Fact-checking with Iterative Retrieval and Verification}

\author{
 \textbf{Zhuohan Xie\textsuperscript{1}} \ 
 \textbf{Rui Xing\textsuperscript{1, 2}} \ 
 \textbf{Yuxia Wang\textsuperscript{1}} \ 
 \textbf{Jiahui Geng\textsuperscript{1}}  \\
 \textbf{Hasan Iqbal\textsuperscript{1}}  \ 
 \textbf{Dhruv Sahnan\textsuperscript{1}} \ 
 \textbf{Iryna Gurevych\textsuperscript{1}} \ 
 \textbf{Preslav Nakov\textsuperscript{1}} \\
 \\
 \textsuperscript{1}MBZUAI,
 \textsuperscript{2}The University of Melbourne \\
\texttt{\{zhuohan.xie, preslav.nakov\}@mbzuai.ac.ae} \\
}

\begin{document}
\maketitle
\begin{abstract}
Fact-checking long-form text is challenging, and it is therefore common practice to break it down into multiple atomic claims. The typical approach to fact-checking these atomic claims involves retrieving a fixed number of pieces of evidence, followed by a verification step. However, this method is usually not cost-effective, as it underutilizes the verification model's internal knowledge of the claim and fails to replicate the iterative reasoning process in human search strategies.
To address these limitations, we propose \ourframework, a novel agent-based framework that integrates evidence retrieval and claim verification in an iterative manner. Specifically, \ourframework employs a unified mechanism to decide whether to provide a final answer or generate a subsequent search query, based on its confidence in the current judgment.
We compare \ourframework with other strong fact-checking frameworks and find that it achieves slightly better performance while reducing large language model (LLM) costs by an average of 7.6 times and search costs by 16.5 times. These results indicate that \ourframework holds promise for application in large-scale fact-checking operations.
Our code is available at \url{https://github.com/mbzuai-nlp/fire.git}.
\end{abstract}
% Why we need to use agents for fact checking?
% Why we choose to do multi modal planning
% Why the benchmark is promptly needed now.

\input{sections/intro}
\input{sections/relatedwork}
\input{sections/framework}

\input{sections/experiments}

\input{sections/results}
\input{sections/discussion}

\section{Conclusions and Future Work}

Conventional fact-checking systems typically separate the steps of evidence retrieval and claim verification, leading to suboptimal utilization of the verification models' internal knowledge. To address this, we propose \ourframework, a novel framework that integrates evidence retrieval and claim verification in an iterative process. \ourframework enables LLMs to leverage their internal knowledge for judgment and only rely on external evidence retrieval when uncertain. Our experiments on multiple datasets demonstrate that \ourframework not only slightly improves accuracy but also reduces LLM computation costs by an average of 7.6 times and search costs by 16.5 times, making it highly efficient for production use.
Additionally, we performed a detailed error analysis, which revealed issues with the benchmarking datasets quality. These findings highlight the need for further research into edge cases, rather than relying solely on automatic metrics for evaluation.

We identify several promising directions for future work, which include:
(1) Integrating memory banks to store verification results, allowing the system to reuse previous results instead of repeatedly executing the entire process;
(2) Expanding the system to support additional modalities, such as code and images; and
(3) Revisiting existing public fact-checking datasets, incorporating personal opinions when addressing ambiguous cases, and adding claims that require rarer and more complex knowledge, where evidence retrieval is essential.

% \paragraph{Expand to more modalities with more tools for calling.}

% \paragraph{Integrate memory bank so that the language model can rely on it for the same or similar claims.}

\section*{Limitations}

We acknowledge several limitations in this work that we plan to address in future research. First, to maintain the efficiency of our framework, we implement the ``Final Answer or Next Search Query'' mechanism in a compact manner, allowing it to retrieve evidence, assess confidence in knowledge, and verify the final answer within a single step. Ideally, this process could be separated to include a standalone confidence estimation step, which would enhance both flexibility and interpretability. We leave this exploration to future work.
Second, to ensure a fair comparison across multiple fact-checking datasets, our system adopts a binary labeling scheme (``True'' or ``False'') and standardizes labels across datasets. However, this approach may not fully capture the complexity of factual labels in real-world settings. We intend to incorporate fine-grained labeling schemes in future research.
Finally, in this study, we rely on SerpAPI with its default settings. While we did not investigate in detail how evidence is retrieved, we believe future work could explore this aspect further to optimize the selection of the most relevant evidence for a given claim.

\section*{Ethical Statement and Broad Impact}

\paragraph{Data License} 
A primary ethical consideration is the data license. We reused pre-existing dataset, \factbench, \factool, \felmwk, \bingcheck, which have been publicly released and approved for research purposes. We adhere to the intended usage of all these dataset licenses.

% Ethical Statement
\paragraph{Ethical Statement}
 We acknowledge that our system relies on LLMs, which can sometimes produce biased or incorrect judgments due to the data used in their pre-training or biases present in external sources. Additionally, there is the risk of over-reliance on the system for making critical factual judgments without human oversight. To mitigate these risks, we strongly encourage human reviewers to be involved in decision-making, especially in high-stakes domains such as legal, political, or medical contexts.

\paragraph{Broad Impact}
\ourframework has the potential to advance the field of automated fact-checking by enhancing its efficiency and accessibility. Its capability to iteratively retrieve evidence while minimizing computational costs will empower a broader range of users—including journalists, researchers, and the general public—to verify factual information with greater ease. Furthermore, \ourframework can be applied to large-scale implementations, such as integration into search engines and social media platforms, thereby contributing to efforts to combat the spread of misinformation.

\section*{Acknowledgments}

We thank our reviewers for their valuable reviews and feedback, which significantly contributed to the improvement of our paper.

\bibliography{custom}

\newpage

\appendix

\input{sections/appendix}

\end{document}

%% file: sections/intro.tex
\section{Introduction}
\label{sec-intro}

\begin{quote}
    \textit{``Every man has a right to his opinion, but no man has a right to be wrong in his facts.''}  - Bernard M. Baruch
\end{quote}

Large language models (LLMs) have demonstrated exceptional performance across a wide range of tasks, including both language comprehension and generation~\citep{llm_survey, xie-etal-2023-next}. Consequently, LLMs are now widely applied in various domains~\citep{DBLP:conf/emnlp/XieLCL23}, and many users increasingly rely on the information they provide. However, this reliance is problematic, as LLMs are capable of producing outputs that are highly confident but factually incorrect, highlighting the critical need for robust fact-checking systems~\citep{mmfcsurvey23}. 
However, fact-checking the entire output of LLMs in a single step is highly challenging. To address this, \citet{factscore23} proposed decomposing the content into multiple atomic claims, each of which can be individually verified. While this approach simplifies the fact-checking process, assessing the veracity of these atomic claims remains complex, especially when many require sourcing evidence from the web. Indeed, identifying the most relevant evidence online is a key challenge in fact-checking pipelines~\citep{factcheckgpt}.

\begin{figure*}[t]
     \centering
     \includegraphics[width=1.0\textwidth]{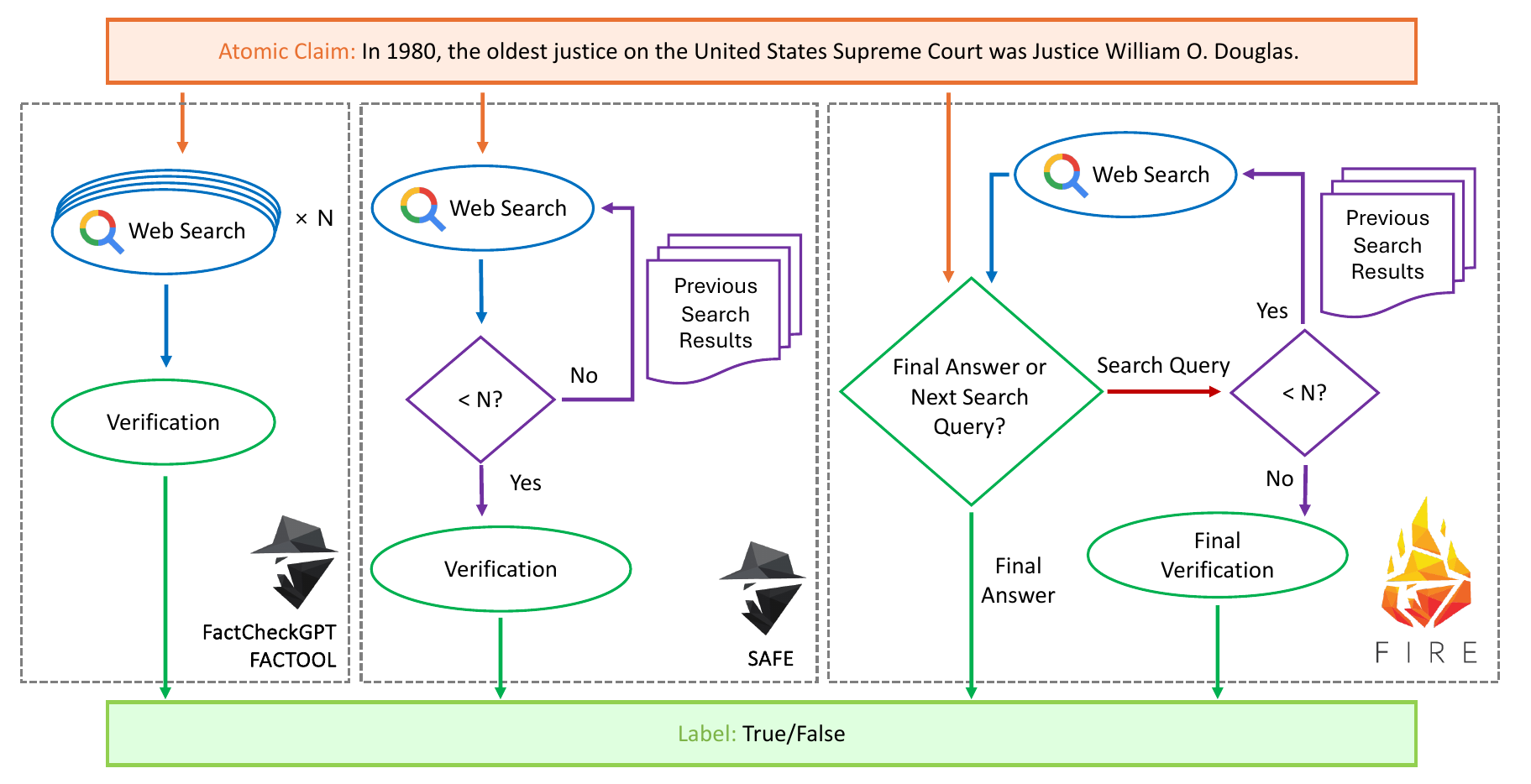}
        \caption{
        \textbf{Comparisons between \ourframework and previous frameworks.} Previous frameworks typically treat web search and claim verification as distinct processes. In contrast, \ourframework integrates interactive retrieval and verification. %
        }
        \label{fig-frameworkcomparison}
\end{figure*}

To address this issue, conventional methods, such as \factool and \factcheckgpt~\citep{factool, factcheckgpt}, frame the problem as a question-answering task, as illustrated on the left side of \autoref{fig-frameworkcomparison}. In these approaches, an LLM is prompted to generate N relevant questions, which are then used as search queries by a web search tool. The search results serve as evidence for LLM to determine the factuality of the claim.
However, we argue that this process is inefficient in two key aspects. First, it underutilizes the internal knowledge already embedded in LLMs during pre-training. For claims involving common knowledge or widely known events, the LLM could confidently assess the claim without relying on external information. Second, generating multiple search queries concurrently does not align with the typical human reasoning process during search~\citep{hu2024avis}. Humans tend to begin with an initial query, gather information, and then refine their perspective on the claim, which often leads to the formulation of more effective follow-up queries.

% Conventional fact-checking methodologies typically encompass multiple stages, including claim detection, evidence retrieval, and factuality assessment.
% All these stages are executed using pre-established pipelines and conducted by LLMs~\citep{factscore23, factool, factcheckgpt, sun2024detecting}.

To address this gap, we introduce \textbf{F}act-checking with \textbf{I}terative \textbf{R}etrieval and V\textbf{E}rification (\ourframework), an innovative agent-based framework that integrates both the internal knowledge of LLMs and external knowledge sources by unifying the verification process and search query generation into a single step.
As illustrated on the right side of \autoref{fig-frameworkcomparison}, \ourframework employs a mechanism to decide whether to produce the final answer or generate a new search query, continuing the evidence-seeking process. This decision is based on the model's confidence in its judgment.
The closest related work to us is \safe~\citep{safe}, depicted in the center of \autoref{fig-frameworkcomparison}. Their method generates web search queries iteratively and subsequently verifies whether the entire retrieved evidence supports the claim. However, this approach lacks flexibility, as it treats evidence retrieval and claim verification as distinct processes, requiring a predetermined fixed number of searches regardless of the claim's complexity. In contrast, our approach integrates evidence retrieval and claim verification into an iterative framework, encouraging the language model to verify based on its own knowledge and conduct searches only when necessary.
\textbf{Our experiments demonstrate that our method significantly reduces the computational costs of LLMs by an average factor of 7.6, as well as search-related costs by a factor of 16.5, all while maintaining fact-checking performance.}

In summary, our contributions are as follows:
\begin{itemize}
    \item We present \ourframework, a simple yet effective interactive framework for fact-checking. Through extensive experiments conducted across multiple datasets, we demonstrate that our framework significantly reduces the LLM computational and search costs, making it a better option for large-scale production.
    \item Our ablation studies demonstrate that the step-by-step reasoning process enhances the model's confidence in fact-checking, particularly with \fouromini. For \fouro, we observed a similar trend; however, the effect was not as pronounced as that seen with \fouromini.
    \item We conducted an error analysis and identified several quality issues in the current benchmark datasets, including the presence of ungrounded claims. Additionally, the strict reasoning capabilities of the LLM may incorrectly classify some debatable claims as non-factual.
\end{itemize}

%% file: sections/relatedwork.tex
\section{Related Work}

% Our work is closely related to fact-checking and the use of large language model agents.

\paragraph{LLM Factuality}

Despite the remarkable capabilities of LLMs~\citep{gpt3,gpt-4, llm_survey}, the auto-regressive learning objective does not inherently offer strong guarantee or enforce the learning of factual accuracy in the training process, making these models produce content that deviates from real-world facts~\citep{wang2024factuality}.
On average, there are 5\%-10\% false claims in responses of GPT-4~\citep{gpt-4} and LLaMA-2~\citep{llama2} on world-knowledge questions~\citep{openfactcheckdemo}.
Retrieval-augmented generation~\citep{realm} and post-generation fact-checking are essential for ensuring accurate knowledge dissemination. Retrieving highly relevant information plays a pivotal role in both guiding generation as a reference and determining verification results in fact-checking systems~\citep{factcheckgpt}.

The retriever and verifier are the most resource-consuming components in fact-checking systems, in terms of time and cost. Even with the inexpensive APIs (e.g., Serper at 0.001 USD per request and GPT-3.5-turbo for verification), verifying an atomic claim costs approximately 0.02 USD, making extensive verification impractical for general users~\citep{openfactcheckdemo}. This high cost limits the ability to verify large volumes of LLM responses, potentially contributing to the spread of misinformation.
Our framework aims to minimize the costs in these two steps, enabling affordable verification for general users. This allows them to easily verify suspicious or doubtful information, enhancing the dissemination of factual information.

\paragraph{Fact Checking with Agents}

The recent advancements in LLMs have spurred significant research on LLM-powered agents, which are capable of reasoning about their environment and making decisions by either invoking external tools or performing internal actions~\citep{toolsurvey24}. These agent frameworks typically consist of several components, including reasoning, tool usage, memory, and multi-agent debate~\citep{agentsurvey24}, many of which can be seamlessly integrated into fact-checking pipelines to enhance the performance of traditional fact-checking systems.
For example, recent works have endowed systems with the ability to call external tools, such as search engines~\citep{factool, factcheckgpt, safe, haluagent24}, recognizing that many claims in the field require additional information for verification. During the verification stage, \citet{multiagentdebatehallucination24} proposed a Markov Chain-based multi-agent debate approach to ensure more rigorous verification by enabling collaborative decision-making among agents based on retrieved evidence.
Our work differs from previous approaches by combining the evidence retrieval and verification stages, leveraging agents' reasoning and tool-use capabilities to more closely simulate human cognitive processes in fact-checking. 

%% file: sections/framework.tex
\section{Framework}

% \subsection{Background}
% Large language models (LLMs) are commonly used to generate extended texts, yet these outputs frequently contain factual inaccuracies.
Assessing the factual accuracy of long-form text presents significant challenges~\citep{factscore23}. To address this, prior approaches have broken down the text into individual checkworthy claims~\citep{factool}. These sentences, referred to as atomic claims, are fact-checked individually, with their factuality scores aggregated to evaluate the overall factual accuracy of the original text.
Previous research indicates that verifying the factuality of atomic claims is the most challenging step in this process~\citep{factcheckgpt}. \textbf{Our work therefore focuses on this critical task: determining the factual accuracy of individual atomic claims, classifying each as either \textit{True} or \textit{False}.}

\subsection{\ourframework}
We present \ourframework, a simple yet effective agent-based framework for interactive claim verification through web searches. As illustrated in \autoref{fig-frameworkcomparison}, \ourframework takes an atomic claim as input and outputs a binary label indicating whether the claim is factual or non-factual. The framework consists of three key components:
\textit{Final Answer or Next Search Query}, \textit{Web Search}, and \textit{Final Verification}, each of which we will explain below.

\paragraph{Final Answer or Next Search Query}
% Previous approaches typically separate the evidence retrieval process from verification. Specifically, they pre-define a fixed number of evidence pieces and then conduct web searches either concurrently (e.g., \factool and \factcheckgpt in \autoref{fig-frameworkcomparison}) or sequentially (e.g., \safe in \autoref{fig-frameworkcomparison}).
% As outlined in \autoref{sec-intro}, the traditional rigid approach lacks adaptability. In contrast,
We introduce a unified method, \textit{Final Answer or Next Search Query} $f(\cdot)$, which integrates claim verification with search query generation. Given an atomic claim $c$, this component decides whether to produce a final answer $a$ or generate an additional search query $q$. This decision is guided by both an external evidence set $E$, derived from search results, and the internal knowledge $k$ of the language model, acquired during pre-training.
At the outset, no evidence has been retrieved, meaning that the evidence set $E$ is initially empty. Consequently, the decision relies solely on the internal knowledge $k$. As shown in \autoref{equ-finalanswerornext}, we incorporate confidence estimation into the reasoning process to determine the next action. If the model's confidence is sufficiently high, it outputs a final answer $a$; otherwise, it generates an additional query $q$.

\begin{equation}
\label{equ-finalanswerornext}
    f(c, E, k) = 
\begin{cases} 
a, & \text{ if confident} \\
q, & \text{ if not confident}
\end{cases}
\end{equation}

This method offers greater flexibility by eliminating the need to retrieve a fixed number of evidence items before verification, thereby largely reducing search costs. A detailed description of the prompt used for this component is provided in \autoref{appendix-verificationprompt}.

% \end{equation}

% Previous approaches typically include a verification model as the final step, where the model evaluates the entire set of evidence to make a judgment. However, as discussed in  In response, we propose a verification method with uncertainty estimation. Our approach takes the original atomic claim and all previously retrieved search results as input, requiring the model to make a judgment based on the current state of the evidence. If the model is confident in its decision, it should produce the final label; otherwise, it generates a query aimed at  

\paragraph{Web search}
When the language model determines that a web search is necessary and issues a search query $q$, we retrieve results using Google Search via the SerpAPI\footnote{\url{https://serpapi.com}}, following prior work~\citep{safe}. This API returns the retrieved snippets as a single string, which we use as the new evidence $e$. We then append $e$ to the existing evidence set $E$ to form the updated evidence set $E'$ for the next iteration, as shown in \autoref{equ-nextsearch}. 

\begin{equation}
\label{equ-nextsearch}
    E' = E \cup e, e = \text{Search}(q) 
\end{equation}

% We utilize  interface. Given a query, it returns the search results in string format.

\paragraph{Final Verification}
Due to the inherent difficulty of confidently verifying certain claims, even with supplementary evidence, we impose an upper limit on the number of retrieval steps. As shown in \autoref{equ-finalverification}, once this limit is reached, the model performs a final verification $f'(\cdot)$ based on all previously retrieved evidence. The detailed prompt for this process is provided in \autoref{appendix-finaldecision}.

\begin{equation}
\label{equ-finalverification}
\begin{cases} 
a = f'(c, E, k), &  n \geq N \\
e = \text{Search}(q), & n < N 
\end{cases}
\end{equation}

\newblock

% In our preliminary experiments, we identified two key issues: (1) During sequential search query generation, language models often produce repetitive queries, even when explicitly instructed to generate new queries that retrieve information relevant to the claim. (2) LLMs tend to exhibit overconfidence in their reasoning, sometimes resulting in no search being performed, despite lacking sufficient information to make a judgment on the claim.

% We explore methods for addressing query repetition in \autoref{sec-searchquerygeneration} and overconfidence in \autoref{uncertainty_methods}.

% In our preliminary studies, we have identified that there exists an issue regarding sequentially search query generation with language models, where they tend to generate repetitive search queries even though it is prompted to generate queries that are supposed to obtain new information regarding the claim. We present the analysis in \autoref{appendix-searchquery}. To mitigate such issues, we explore methods to generate search queries.

\subsection{Prevention of Repetitive Search Queries}
\label{sec-searchquerygeneration}
In our preliminary studies, we identified a recurring issue with sequential search query generation using language models: the tendency of these models to generate repetitive queries. This occurs even when the models are explicitly instructed to generate queries targeting new, claim-relevant information. As a result, identical queries are repeatedly submitted to web search tools, leading to inefficient use of search resources. To address this issue, we investigate following methods for enhancing search query generation and reducing repetition.

\paragraph{Early Termination} The iterative process is terminated when consecutive queries or retrieved results exhibit a high degree of similarity, indicating diminishing returns.

\paragraph{Diversity Prompt} We introduce additional prompts to encourage the model to generate more diverse queries when consecutive similar queries or search results are detected.

% indicate that recent queries or retrievals show significant similarity. The model is then prompted to generate more diverse queries.

% The first approach is early termination, where the retrieval process halts once  The second approach involves actively prompting the model to generate more diverse queries.  These strategies aim to enhance the efficiency of retrieval-augmented generation (RAG) systems.

\subsection{Prevention of Verification Overconfidence}
\label{uncertainty_methods}
LLMs can exhibit strong calibration abilities across diverse tasks~\citep{kadavath2022language,geng2024survey}. Consequently, they are aware of their confidence levels during the claim verification process. However, our preliminary analysis reveals that LLMs often demonstrate excessive strictness and unwarranted confidence in certain cases, leading to errors. Considering this, we explore several techniques to prevent overconfidence in verification:

% \paragraph{Reason} We require models to output reasons/explanations to backup their decisions. This is our default option.

% \paragraph{No Reason} Requiring models to provide reasons or explanations increases token consumption, which raises the overall cost and slows down inference speed. To improve efficiency, we opted for a setting to output only the label.

% RX: merge At Least One and At Least Two
\paragraph{At Least One/Two} At Least One requires models to retrieve at least one evidence during the verification, which increase the probability of eliminating overconfidence. Similarly, we also adopted a more aggressive approach At Least Two  to retrieve a second evidence to reduce the uncertainty.

\paragraph{Inclusive Prompt} In this setting, we prompt models to be ``less strict, open-minded and avoid being over confident'' to encourage models to reflect on their confidence level of answers.

%% file: sections/experiments.tex
\section{Experiments Setup}

% Our framework is designed to enhance the retrieval of evidence only when necessary and to generate a binary label, either True or False, for a given claim. We begin with the atomic claim. The processes of decomposing entire LLM responses into multiple claims or determining the checkworthiness of claims fall outside the scope of this work.

\subsection{Datasets}
In our study, we utilized four datasets from prior research that align with our experimental setup: FacTool~\citep{factool}, FELM~\citep{felm}, \factcheckbench~\citep{factcheckgpt}, and \bingcheck~\citep{selfchecker}.
FacTool and FELM provide factuality claims across multiple domains. From these, we selected instances requiring world knowledge for verification, which we refer to as \factoolqa and \felmwk, both annotated with binary labels (\textit{True} or \textit{False}). Our selection was motivated by the need to focus on claims that challenge models to use external knowledge, a critical aspect of factual verification.

For \factcheckbench and \bingcheck, we consolidated the original four-label classification (\textit{supported}, \textit{partially supported}, \textit{not supported}, \textit{refuted}) into a binary format by merging \textit{supported} and \textit{partially supported} into \textit{True}, treating \textit{refuted} as \textit{False}, and excluding \textit{not supported}. This binarization aligns these datasets with the others and simplifies evaluation, focusing on clear-cut factuality decisions.
We sampled a subset of \bingcheck due to its class imbalance (3,581 \textit{True} claims versus 42 \textit{False} claims), selecting 100 \textit{True} claims for our test set. This sampling was essential to create a more balanced and manageable test set, ensuring that evaluation metrics reflect performance on both classes without being dominated by the majority class.
In \felmwk, we retained un-split claims to maintain contextual integrity, which is crucial for accurate verification. Full dataset statistics are provided in \autoref{tab-dataset-stats}.

\begin{table}[t]
\small
\centering
\begin{adjustbox}{max width=\columnwidth}
\begin{tabular}{l|rrr}
\toprule
\textbf{Dataset} & \textbf{\#True} & \textbf{\#False}  & \textbf{Total} \\
\midrule
\factcheckbench & 472 & 159 & 631 \\
\factoolqa & 177 & 56  & 233 \\
\felmwk & 99 & 85 &  184 \\
\bingcheck & 100 & 42 & 142 \\
% \midrule
% HaluEval & 3,692 & 815 & 4,507 \\
\bottomrule
\end{tabular}
\end{adjustbox}
\caption{\textbf{Statistics of the datasets after processing.}}
\label{tab-dataset-stats}
\end{table}

% FacTool and FELM both contain factuality claims across multiple domains. From these datasets, we selected instances that require world knowledge for verification, which we refer to as \factoolqa and \felmwk. These two datasets provide binary annotations by human annotators, labeled as either \textit{True} or \textit{False}. In contrast, \factcheckbench and \bingcheck address more nuanced scenarios, framing the task as a four-label classification problem with the categories: \textit{supported}, \textit{partially supported}, \textit{not supported}, and \textit{refuted}.
% To ensure consistency across datasets, we adopted the approach of \citet{openfactcheckdemo}, merging the \textit{supported} and \textit{partially supported} labels into a single category, \textit{True}, and treating the \textit{refuted} label as \textit{False} while excluding the \textit{not supported} label from further analysis. Additionally, as the original dataset for \bingcheck is highly imbalanced—containing 3,581 \textit{True} claims but only 42 \textit{False} claims after merging—we sampled 100 \textit{True} claims to construct our test set.
% For the \felmwk dataset, the original data includes both split (532 claims) and un-split claims (184), all annotated with labels. Split claims are essentially longer claims divided using a “.” delimiter. We decided to use the un-split long claims because the split versions often miss necessary context, resulting in ambiguous interpretations. The statistics for the processed datasets are provided in \autoref{tab-dataset-stats}.

In our experiments, we first use the Factcheck-Bench dataset as a development set to optimize the settings for our framework. We then evaluate its performance on the remaining three datasets, comparing it with other competitive fact-checking systems. 

\subsection{Language Models}
We investigate several state-of-the-art (SOTA) language models, including proprietary models from two prominent families: GPT models~\citep{gpt-4o, gpt-o1} and Claude models~\citep{claude3}, as detailed in \autoref{tab-lms}. In addition, we assess two open-source models: \llama3.1-Inst 8B~\citep{llama3.1} and \mistral-Inst 7B~\citep{mistral-7b}.

\begin{table}[t]
\small
\centering
\begin{tabular}{lL{5cm}}
\toprule
\textbf{Family} & \textbf{Name} \\
\midrule
GPT & \fouro, \fouromini, \oone, \oonemini \\ 
Claude & \haiku, \opus, \sonnet \\ 
\midrule
\llama & \llama3.1-Inst 8B \\
\mistral & \mistral-Inst 7B \\
\bottomrule
\end{tabular}
\caption{\textbf{Model families and specific model names used in this study.}}
\label{tab-lms}
\end{table}

\subsection{Compared Fact-checking Frameworks}

We select several SOTA fact-checking frameworks for comparison. Additionally, we introduce two baseline models: Random and Always True/False. To further assess the impact of LLM reasoning and evidence retrieval in fact-checking, we include two ablation settings: \ourframework (No Reason) and \ourframework (No Search).

\paragraph{\factool} is adaptable across domains and tasks, using a tool-augmented framework that integrates external tools like Google Search and Python interpreters to assess the factuality of content from large language models. However, this can introduce complexity and depend on the accuracy of these external tools.

\paragraph{\factcheckgpt} excels in fine-grained factuality evaluation through a detailed benchmark with annotations at the claim, sentence, and document levels. While resource-intensive, it provides valuable insights into specific stages of factual inaccuracies.

\paragraph{\safe} uses a search-augmented approach to verify long-form content by breaking it down into individual facts and checking them via Google Search. This method is cost-effective compared to human annotation but depends on the reliability of search engine results, which can vary and introduce biases.

\paragraph{Random} assigns the predicted label for each claim in the test set randomly, choosing between \textit{True} and \textit{False} with equal probability.

\paragraph{Always True/False} is an approach that always predicts a single label -- either \textit{True} or \textit{False} -- for all claims in the test set.

\paragraph{\ourframework (No Reason)} utilizes the same framework as \ourframework; however, it is explicitly instructed not to articulate its reasoning process in the output. This modification aims to assess the impact of explicitly stating the step-by-step reasoning process on the results.

\paragraph{\ourframework (No Search)} employs the same framework as \ourframework; however, it is not permitted to invoke the search tool. This configuration is designed to evaluate the model's ability to perform fact-checking without retrieving any supporting evidence.

% , and we demonstrate their statistics in \autoref{tab-dataset-stats}. 

% use four datasets from \citet{openfactcheck} for this work.
% More specifically, we use Factcheck-Bench for developing and found our best setting and we then test the generalization ability across three datasets.

% Some studies~\citep{selfchecker}  We present their dataset statistics in \autoref{tab-dataset-stats-origin}.

\subsection{Evaluation Metrics}
In this work, we investigate the trade-off between computational cost and fact-checking performance.

\paragraph{Performance} We evaluate precision, recall, and F1 scores for both positive and negative classes.

\paragraph{Computational Cost} We report the financial costs of LLM API calls for proprietary models and GPU rental expenses for open-source models, alongside an analysis of API costs from search engine queries and a breakdown of the total time spent on the fact-checking process. The experiments using open-source models were conducted on an NVIDIA RTX 6000 GPU at an estimated cost of \$0.79 per hour, while search queries via SerpAPI incurred approximately \$0.00105 per search. 

% Think how do we evaluate the intermediate steps.
% Can we use approaches from OpenFactCheck leaderboard on this?

%% file: sections/results.tex
\section{Results}
\label{sec-results}

In this section, we first present preliminary studies on \factcheckbench (\autoref{subsec-preliminarystudies}), focusing on three key aspects: language models, prevention of repetitive search queries, and prevention of verification overconfidence. These studies aim to identify the most appropriate configurations for our framework. Subsequently, we compare \ourframework to other strong fact-checking frameworks across three additional datasets (\autoref{subsec-comparsiontosota}) to evaluate the generalization capabilities of our approach.

\subsection{Preliminary studies}
\label{subsec-preliminarystudies}

\paragraph{Language Models}

We present a performance comparison of various language models in \autoref{tab-languagemodels}. Overall, proprietary language models generally outperform open-source models, likely due to their larger size and more sophisticated training in reasoning and tool utilization. Among the proprietary models, the latest and most advanced offerings from different organizations—specifically \oone from OpenAI and \sonnet from Anthropic—exhibit the best performance. Although the more economical model, \fouromini, performs slightly worse than the top-performing \oone, it offers a cost savings of 766 times. This suggests that for fact-checking tasks, the most advanced models may not be necessary; \fouromini can serve as a sufficiently capable alternative at a significantly lower cost. We will continue our preliminary studies using \fouromini.
% Additionally, since the latter models are trained on more up-to-date datasets, they may contain knowledge that is more relevant for fact-checking tasks.

% Furthermore, we present the computational cost associated with different language models in \autoref{tab-languagemodels}. It is evident that \fouromini incurs the lowest cost, both in terms of LLM calls and search operations, while also being the fastest model. 

% To explore this further, we provide the distribution of search calls made by each model in \autoref{fig-searchnumbers}. Notably, \fouromini demonstrates high confidence, relying primarily on its internal knowledge to verify claims without initiating additional search requests.
% Given that the performance of \fouromini is relatively close to the best models while having the lowest cost, we consider it the optimal choice for large-scale production. As a result, we continue our experiments using \fouromini.

\setlength{\tabcolsep}{3pt}
\begin{table}[t]
\centering
\resizebox{\columnwidth}{!}{
\begin{tabular}{lccccccc}
\toprule
 \multirow{2}{*}{\textbf{LLM}}  & \multirow{2}{*}{\textbf{\shortstack{LLM+Search\\ Cost (\$)}}} & \multicolumn{3}{c}{\textbf{Label = True}} & \multicolumn{3}{c}{\textbf{Label = False}} \\ &
 & \textbf{Prec} & \textbf{Recall} & \textbf{F1} & \textbf{Prec} & \textbf{Recall} & \textbf{F1} \\ 
 \midrule
 \fouromini &  \textbf{0.19}+\textbf{0.44} & 0.91 & 0.84 & 0.87 & 0.61 & 0.74 & 0.67 \\
 \fouro & 10.45+1.47 & 0.92 & 0.79 & 0.85 & 0.56 & 0.79 & 0.66 \\
\oone & 145.66+0.80 & 0.91 & \textbf{0.86} & \textbf{0.88} & \textbf{0.64} & 0.75 & 0.69 \\
\oonemini & 20.06+1.13 & 0.89 & 0.81 & 0.85 & 0.56 & 0.71 & 0.62 \\
\haiku & 0.56+0.85 & 0.9 & 0.81 & 0.85 & 0.56 & 0.73 & 0.64 \\
\opus & 48.64+1.43 & 0.92 & 0.81 & 0.86 & 0.58 & 0.79 & 0.67 \\
\sonnet & 13.21+1.63 & \textbf{0.94} & 0.79 & 0.86 & 0.58 & \textbf{0.85} & \textbf{0.69} \\
\midrule
\llama3.1-Inst 8B & 3.95+2.27 & 0.89 & 0.74 & 0.8 & 0.48 & 0.72 & 0.57 \\
 \mistral-Inst 7B & 1.84+1.22 & 0.85 & 0.67 & 0.75 & 0.4 & 0.66 & 0.5 \\
\bottomrule
\end{tabular}}
\caption{\textbf{\textbf{Fact-checking performance and cost comparisons between different language models within \ourframework on \factcheckbench.}}}
\label{tab-languagemodels}
\end{table}

\paragraph{Prevention of Repetitive Search Queries}
\label{subsec-searchgenmethods}

We conducted an experimental analysis to evaluate the impact of \textbf{Early Termination} and \textbf{Diversity Prompt} on mitigating the generation of repetitive search queries. To assess query similarity, we employed Sentence-BERT (all-MiniLM-L6-v2; \citet{reimers-gurevych-2019-sentence}) with a similarity threshold of 0.9, as established by \citet{shashavali2019sentence}.
\autoref{tab-diverse} presents experimental results, where \textbf{window size} refers to the predefined number of consecutive similar queries or retrieval results. Once this threshold is reached, early termination is triggered to prevent further query generation and retrieval. If the model generates queries or retrieves results exhibiting high similarity within this window, the system also activates an early stopping mechanism. The results indicate that optimizing the similarity window size effectively reduces search costs without compromising the model's performance. However, our findings suggest that the diversity prompt does not enhance performance.
In our optimal configuration, we selected a window size of 2 without utilizing the diversity prompt.
%RX: is similarity threshold more appropriate?
\setlength{\tabcolsep}{3pt}

\begin{table}[t]
\centering
\resizebox{\columnwidth}{!}{
\begin{tabular}{ccccccccc}
\toprule
 \multirow{2}{*}{\shortstack{\textbf{Window}\\ \textbf{Size}}}  & 
 \multirow{2}{*}{\shortstack{\textbf{Diversity}\\ \textbf{Prompt}}} &
 \multirow{2}{*}{\textbf{\shortstack{LLM+Search\\ Cost (\$)}}} & \multicolumn{3}{c}{\textbf{Label = True}} & \multicolumn{3}{c}{\textbf{Label = False}} \\ &
 & & \textbf{Prec} & \textbf{Recall} & \textbf{F1} & \textbf{Prec} & \textbf{Recall} & \textbf{F1} \\ 
 \midrule
\multirow{2}{*}{2} & \xmark & 0.17+0.29 & \textbf{0.92} & 0.83 & \textbf{0.87} & \textbf{0.61} & \textbf{0.77} & \textbf{0.68} \\
                   & \cmark & 0.16+0.29 & 0.91 & 0.81 & 0.86 & 0.57 & 0.76 & 0.65 \\
\arrayrulecolor{white}
\midrule
\multirow{2}{*}{3} & \xmark & 0.17+0.36 & 0.91 & 0.82 & \textbf{0.87} & 0.60 & 0.77 & 0.67 \\
                   & \cmark & 0.18+0.36 & 0.91 & 0.82 & 0.86 & 0.59 & 0.76 & 0.66 \\
\midrule
\multirow{2}{*}{4} & \xmark & 0.18+0.39 & 0.91 & 0.81 & 0.86 & 0.57 & 0.76 & 0.65 \\
                   & \cmark & 0.18+0.39 & 0.91 & 0.82 & 0.86 & 0.59 & 0.76 & 0.66 \\ 
\arrayrulecolor{black}
\midrule
Default & - & 0.19+0.44 & 0.91 & \textbf{0.84} & \textbf{0.87} & \textbf{0.61} & 0.74 & 0.67 \\
\bottomrule
\end{tabular}}
\caption{\textbf{\ourframework performance across various window sizes, with and without the use of prompts for generating diverse queries on \factcheckbench.}}
\label{tab-diverse}
\end{table}

\paragraph{Prevention of Verification Overconfidence}
\label{subsec-uncertaintymethods}
We present the performance and cost of various overconfidence prevention approaches for verification on \factcheckbench in \autoref{tab-uncertainty}. Interestingly, the \textbf{At Least One/Two} settings, which aggressively retrieve additional evidence, result in higher search costs without improving fact-checking performance compared to the \textbf{Default} setting, where no explicit constraints are placed on web search. This supports our hypothesis that most atomic claims are relatively straightforward and do not require extensive external web searches for verification. In fact, introducing additional searches may introduce noise, negatively impacting performance. The \textbf{Inclusive} setting encourages models to be more flexible and open to alternative interpretations of evidence, which reduces the need for queries but also leads to lower overall performance. Based on these observations, we maintain the \textbf{Default} setting, leveraging the language model's reasoning capabilities without imposing additional search constraints.

% Overall, we notice that ,  achieves best performance among most metrics against other methods with minor improvement. However, except for \textbf{Inclusive}, \textbf{Reason} setting largely surpasses other methods in terms of Cost, which indicates outputting reasons of their decisions can both improve their accuracy in verification and reduce the need for evidences, which naturally fits our framework. In addition, we also found \textbf{No-reason}, \textbf{At least one} and \textbf{At least two} aggressively retrieve evidences while not increasing too much of the performance. 

% We also present the results for \fouromini failed cases in \autoref{tab-failedcases}. Interestingly, \textbf{At least two} achieved the best performance among all methods, followed by \textbf{At least one} and \textbf{No reason}. These methods incurred five times the cost in searching for evidence compared to the others, suggesting that additional evidence is crucial for these failed cases. However, their performance declined  on the full \factcheckbench, making them less effective in general cases.

\setlength{\tabcolsep}{3pt}
\begin{table}[t]
\centering
\resizebox{\columnwidth}{!}{
\begin{tabular}{lccccccc}
\toprule
 \multirow{2}{*}{\textbf{Approach}}  & \multirow{2}{*}{\textbf{\shortstack{LLM+Search\\ Cost (\$)}}} & \multicolumn{3}{c}{\textbf{Label = True}} & \multicolumn{3}{c}{\textbf{Label = False}} \\ &
 & \textbf{Prec} & \textbf{Recall} & \textbf{F1} & \textbf{Prec} & \textbf{Recall} & \textbf{F1} \\ 
 \midrule
% Reason (default)  & 0.19+0.44 & 0.91 & \textbf{0.84} & \textbf{0.87} & \textbf{0.61} & 0.74 & \textbf{0.67} \\
% No reason  & 0.11+0.94 & 0.91 & 0.83 & 0.87 & 0.59 & 0.74 & 0.66 \\
At Least One  & 0.21+0.83 & \textbf{0.92} & 0.81 & 0.86 & 0.58 & \textbf{0.78} & 0.67 \\
At Least Two  & 0.22+0.87 & 0.91 & 0.79 & 0.84 & 0.55 & 0.77 & 0.64 \\
Inclusive  & 0.20+\textbf{0.42} & 0.91 & 0.81 & 0.86 & 0.58 & 0.77 & 0.66 \\ \midrule
Default  & \textbf{0.19}+0.44 & 0.91 & \textbf{0.84} & \textbf{0.87} & \textbf{0.61} & 0.74 & \textbf{0.67} \\
\bottomrule
\end{tabular}}
\caption{\textbf{\ourframework performance using different verification overconfidence prevention approaches on \factcheckbench.}}
\label{tab-uncertainty}
\end{table}

% \begin{table}[t]
% \centering
% \resizebox{\columnwidth}{!}{
% \begin{tabular}{lccccccc}
% \toprule
%  \multirow{2}{*}{\textbf{Approach}}  & \multirow{2}{*}{\textbf{\shortstack{LLM+Search\\ Cost (\$)}}} & \multicolumn{3}{c}{\textbf{Label = True}} & \multicolumn{3}{c}{\textbf{Label = False}} \\ &
%  & \textbf{Prec} & \textbf{Recall} & \textbf{F1} & \textbf{Prec} & \textbf{Recall} & \textbf{F1} \\ 
%  \midrule
% Reason (default)  & \textbf{0.01+0.02} & 0.23 & 0.19 & 0.21 & 0.1 & 0.13 & 0.12 \\
% No reason  & 0.01+0.09 & 0.39 & 0.34 & 0.37 & 0.22 & 0.26 & 0.24 \\
% At least one  & 0.02+0.06 & 0.42 & 0.31 & 0.36 & 0.29 & 0.39 & 0.33 \\
% At least two & 0.02+0.07 & \textbf{0.46} & \textbf{0.34} & \textbf{0.39} & \textbf{0.32} & \textbf{0.43} & \textbf{0.37} \\
% Inclusive & 0.02+0.02  & 0.28 & 0.22 & 0.25 & 0.17 & 0.22 & 0.19 \\
% \bottomrule
% \end{tabular}}
% \caption{\textbf{\ourframework performance of failed cases on \factcheckbench using different uncertainty prompts.} These are instances that \fouromini failed to predict the correct label.}
% \label{tab-failedcases}
% \end{table}

\setlength{\tabcolsep}{3pt}
\begin{table*}[t]
% \scriptsize
    \centering
    \resizebox{\textwidth}{!}{
    \small
    \begin{tabular}{@{}l|c|ccc|ccc|ccc|ccc|ccc|ccc@{}}
    \toprule
    \multicolumn{1}{c|}{\multirow{3}{*}{\textbf{Framework}}} & \multicolumn{1}{c|}{\multirow{3}{*} {\textbf{LLM}}}  & \multicolumn{6}{c|}{\textbf{\factoolqa}} & \multicolumn{6}{c|}{\textbf{\felmwk}} & \multicolumn{6}{c}{\textbf{\bingcheck}} \\
    &   & \multicolumn{3}{c|}{\textbf{Label = True}} & \multicolumn{3}{c|}{\textbf{Label = False}} & \multicolumn{3}{c|}{\textbf{Label = True}} & \multicolumn{3}{c|}{\textbf{Label = False}} & \multicolumn{3}{c|}{\textbf{Label = True}} & \multicolumn{3}{c}{\textbf{Label = False}} \\
    & & Prec & Recall & F1 & Prec & Recall & F1 & Prec & Recall & F1 & Prec & Recall & F1 & Prec & Recall & F1 & Prec & Recall & F1 \\
    \midrule
    Random & - & 0.81 & 0.47 & 0.59 & 0.28 & 0.64 & 0.39 & 0.75 & 0.49 & 0.59 & 0.30 & 0.57 & 0.39 & 0.77 & 0.67 & 0.72 & 0.40 & 0.52 & 0.45 \\
    Always True & - & 0.76 & 1.0 & 0.86 & 0 & 0 & 0 & 0.72 & 1.0 & 0.84 & 0 & 0 & 0 & 0.70 & 1.0 & 0.83 & 0 & 0 & 0\\
    Always False & - & 0 & 0 & 0 & 0.24 & 1.0 & 0.39 & 0 & 0 & 0 & 0.28 & 1.0 & 0.44 & 0 & 0 & 0 & 0.30 & 1.0 & 0.46\\
    \midrule
    % \factool & GPT-3.5 Turbo & 0.93 & 0.55 & 0.70 & 0.38 & 0.88 & 0.53 & 0.76 & 0.50 & 0.60 & 0.32 & 0.60 & 0.41 & 0.93 & 0.66 & 0.77 & 0.52 & 0.88 & 0.65\\
    \multirow{2}{*}{\factool} & \fouro & 0.88 & 0.81 & 0.84 & 0.52 & 0.66 & 0.58 & 0.69 & 0.53 & 0.60 & 0.57 & 0.73 & 0.64 & 0.86 & 0.57 & 0.68 &  0.43 & 0.79 & 0.56\\
     & \fouromini & \textbf{0.92} & 0.68 & 0.78 & 0.45 & \textbf{0.82} & 0.58 & 0.67 & 0.37 & 0.48  & 0.51 & \textbf{0.78} & 0.62 & \textbf{0.92} & 0.55 & 0.69 & 0.45 & \textbf{0.88} & 0.60 \\
     \arrayrulecolor{white}
     \midrule
    \multirow{2}{*}{\factcheckgpt} & \fouro & 0.90 & 0.79 & 0.84 & 0.52  & 0.71 & 0.60  & 0.67 & 0.68 & 0.67  & 0.61 & 0.61  & 0.61 & 0.85 & 0.70 & 0.77 & 0.50 & 0.71 & 0.59  \\
     & \fouromini & 0.85 & 0.80 & 0.82 & 0.47  & 0.56 & 0.51 & 0.61 & 0.50 & 0.55   & 0.51 & 0.62 & 0.56 & 0.88 & 0.78 & 0.83 & 0.60 & 0.76 & 0.67 \\
     \midrule
    \multirow{2}{*}{\safe} & \fouro & \textbf{0.92} & \textbf{0.88} & \textbf{0.90} & \textbf{0.66} & 0.77 & \textbf{0.71} & \textbf{0.70} & 0.80 & 0.75 & 0.72 & 0.60 & 0.65 & 0.84 & 0.90 & 0.87 & 0.71 & 0.60 & 0.65 \\
     & \fouromini & \textbf{0.92} & 0.82 & 0.87 & 0.58 & 0.79 & 0.67 & 0.61 & 0.76 & 0.68 & 0.61 & 0.44 & 0.51 & 0.86 & 0.81 & 0.84 & 0.60 & 0.69 & 0.64\\
     \midrule
    \multirow{2}{*}{\ourframework} & \fouro & \textbf{0.92} & \textbf{0.88} & \textbf{0.90} & 0.65 & 0.71 & 0.68 & \textbf{0.70} & \textbf{0.86} & \textbf{0.77} & \textbf{0.77} & 0.54 & 0.63 & 0.86 & 0.88 & 0.87 & 0.70 & 0.67 & 0.68 \\
     & \fouromini & 0.87 & \textbf{0.88} & 0.87 & 0.60 & 0.59 & 0.59 & 0.63 & 0.82 & 0.71 & 0.67 & 0.44 & 0.53 & 0.87 & \textbf{0.91} & \textbf{0.88} & 0.74 & 0.67 & 0.70 \\ 
     \arrayrulecolor{black}
     \midrule
     \multirow{2}{*}{\ourframework (No Reason)} & \fouro & 0.88 & 0.86 & 0.87 & 0.60 & 0.64 & 0.62 & \textbf{0.70} & 0.85 & \textbf{0.77} & \textbf{0.77} & 0.58 & \textbf{0.66} & 0.85 & 0.89 & 0.87 & 0.70 & 0.62 & 0.66 \\
     & \fouromini & 0.87 & 0.84 & 0.86 & 0.55 & 0.61 & 0.58 & 0.65 & 0.84 & 0.73 & 0.71 & 0.47 & 0.57 & 0.84 & 0.87 & 0.85 & 0.66 & 0.6 & 0.62 \\
     \arrayrulecolor{white}
     \midrule
     \multirow{2}{*}{\ourframework (No Search)} & \fouro & 0.86 & 0.87 & 0.88 & 0.61 & 0.54 & 0.57 & 0.69 & \textbf{0.86} & \textbf{0.77} & \textbf{0.77} & 0.55 & 0.65 & 0.86 & \textbf{0.91} & \textbf{0.88} & \textbf{0.79} & 0.64 & \textbf{0.71} \\
     & \fouromini & 0.84 & 0.84 & 0.84 & 0.49 & 0.48 & 0.49 & 0.61 &\textbf{0.86} & 0.72 & 0.7 & 0.36 & 0.48 & 0.83 & 0.9 & 0.87 & 0.71 & 0.57 & 0.63 \\
     \arrayrulecolor{black}
    \bottomrule
    \end{tabular}
    }
    \caption{\textbf{Performance comparisons between different frameworks across multiple datasets.}}
    \label{tab-verification}
    % \vspace{-1em}
\end{table*}

\begin{table}[t]
\centering
\resizebox{\columnwidth}{!}{
\begin{tabular}{lcccccc}
\toprule
 \textbf{Framework} & \textbf{LLM} & \textbf{LLM } & \textbf{Search } & \textbf{Time} \\
 \midrule
 \multirow{2}{*}{\factool} & \fouro & 24.76 & 3.67 & 2.92\\
  & \fouromini & 1.49 & 3.67 & 2.34\\
 \arrayrulecolor{white}
 \midrule
 \multirow{2}{*}{\factcheckgpt} & \fouro & 21.41 & - & 4.25 \\
   & \fouromini & 1.28 & - & 4.09 \\
 \midrule
\multirow{2}{*}{\safe} & \fouro & 6.34 & 2.93 & 4.62\\
 & \fouromini & 0.43 & 2.93 & 4.25 \\
\midrule
\multirow{2}{*}{\ourframework} & \fouro & 3.35 & 0.60 & 1.31 \\
 & \fouromini & 0.14 & \textbf{0.20} & 1.25 \\ \midrule
\arrayrulecolor{black}
\midrule
\multirow{2}{*}{\ourframework (No Reason)} & \fouro & 1.65 & 0.68 & 0.57 \\
 & \fouromini & \textbf{0.07} & 0.59 & \textbf{0.54} \\
 \arrayrulecolor{white}
 \midrule
 \multirow{2}{*}{\ourframework (No Search)} & \fouro & 1.70 & - & 1.03 \\
 & \fouromini & 0.11 & - & 1.34 \\
 \arrayrulecolor{black}
\bottomrule
\end{tabular}}
\caption{\textbf{LLM/Search cost (USD) and time (hrs) for evaluating the total 559 atomic claims in \factoolqa, \felmwk, and \bingcheck.} We use SerperAPI for \factool, \safe and \ourframework for search, while \factcheckgpt has its own implemented scrapping technique.}
\label{tab-frameworkcost}
\end{table}

\subsection{Comparisons to Other Frameworks}
\label{subsec-comparsiontosota}

We present a performance comparison of our framework against other frameworks in \autoref{tab-verification} and a cost analysis in \autoref{tab-frameworkcost}. As shown, all frameworks exhibit similar performance, with a small gap of approximately 0.2. Our framework, using \fouro, performs slightly better, achieving superior results on 7 out of 18 metrics, followed closely by \safe with \fouro at 6 metrics. This suggests that all frameworks can effectively perform fact-checking for most claims, although they may encounter difficulties with challenging examples, which we analyze further in \autoref{sec-erroranalysis}.
Regarding the necessity of evidence retrieval in fact-checking, we observe a relatively larger performance drop in \factool when evidence search is omitted, compared to a smaller drop in \felmwk and \bingcheck. This suggests that \factoolqa comprises more rare knowledge than \fouromini, whereas \felmwk and \bingcheck may rely predominantly on common knowledge, for which evidence retrieval is less impactful. Overall, both \fouro and \fouromini perform reasonably well on popular public datasets, highlighting the need for datasets that incorporate more complex claims.
In terms of model size, \fouro generally outperforms \fouromini across most frameworks, indicating that larger models are more effective in detecting misinformation. However, the performance improvement is limited, and the associated costs result in an average increase of 16.7 times in LLM expenses and a three-fold increase in search costs when using \ourframework. Therefore, we argue that cheaper models, such as \fouromini, are a viable option for performing fact-checking tasks.
Furthermore, when considering all frameworks with \fouromini, \ourframework achieves additional cost savings, reducing LLM expenses by 7.6 times and search costs by 16.5 times compared to other frameworks. Thus, we contend that \ourframework, when paired with \fouromini, offers a compelling solution for the large-scale deployment of fact-checking systems.
% In terms of the impact of evidence retrieval, we observe that 

\begin{figure}[t]
     \centering
     \includegraphics[width=\columnwidth]{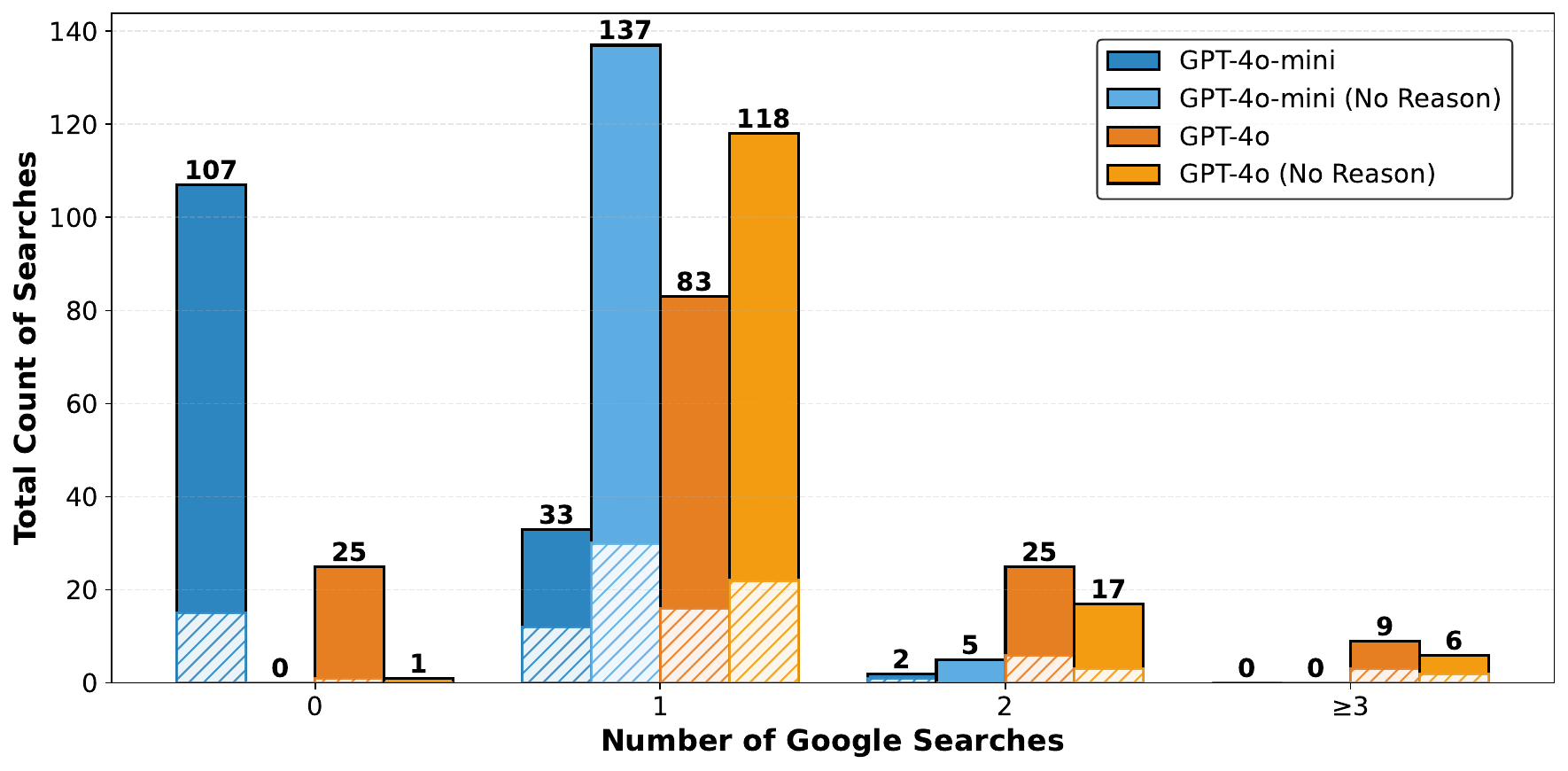}
        \caption{
        \textbf{The effect of reasoning on the number of searches using \fouro and \fouromini within \ourframework on \bingcheck.}
        The shaded area indicates the number of misclassified cases. 
        The x-axis shows the number of web searches, while the y-axis denotes the number of instances.
        }
        \label{fig-bingchecksearchnumbers}
\end{figure}

\autoref{fig-bingchecksearchnumbers} illustrates the impact of reasoning on the number of web searches conducted by \fouro and \fouromini tested on \bingcheck. Notably, \fouromini demonstrates a high level of confidence in making verifications when it is allowed to articulate its reasoning process, resulting in the majority of judgments being made without any searches. Conversely, when not permitted to express its reasoning, there is a significant decrease in the number of instances with zero searches; most cases now involve at least one search, indicating a marked reduction in \fouromini's confidence in its judgments. This observation aligns with previous findings that the presence of CoT reasoning correlates with increased confidence in the model's answers~\citep{cotreasoningconfidence}. While \fouro also shows a decline in confidence when it is not allowed to search, the decrease is less pronounced than that observed in \fouromini.

By combining the performance and cost results presented in \autoref{tab-verification} and \autoref{tab-frameworkcost}, we find that, in the absence of a reasoning process, the costs associated with LLMs can be reduced through fewer completion tokens. However, this reduction leads to increased search costs, resulting in overall performance that is inferior to scenarios in which the models are permitted to engage in step-by-step reasoning. Furthermore, the step-by-step reasoning approach facilitates more effective error analysis.

%% file: sections/discussion.tex
\begin{table*}[t!]
\centering
\resizebox{\textwidth}{!}{\small
\begin{tabular}{l  p{12cm}  ccc  c}
\toprule
% \textbf{Error Type} & \textbf{Description and Examples} & \textbf{\felmwk} & \textbf{\factoolqa} & \textbf{\bingcheck} & \textbf{\# Total} \\
\textbf{Major Issue} & \textbf{Error Type Description} & \textbf{FELM} & \textbf{FacTool} & \textbf{BingCheck} & \textbf{\#Total} \\
    \midrule 
    \multirow{5}{3cm}{I. Dataset Issue} &
    \textbf{1. Not a claim}, e.g.\ a claim only has a name \textit{Elvis Presley.}  & 12 &  1 &  0 & 13 \\
    & \textbf{2. Unclear, ambiguous or subjective claim} e.g.\ there is no record of how many sons \textbf{he} had. & 11 &  7 &  2 & 20 \\
    & \textbf{3. False gold labels}, i.e., the original annotated label might be wrong. For example, claim \textit{choosing organic and local foods that are in season can reduce emissions from transporting food from far away} is labeled as false & 10 &  0 &  1 & 11 \\
    \midrule
    \multirow{8}{3cm}{II. Knowledge Issue} &
    \textbf{4. Complicated science domain expert knowledge} is needed to judge, like astronomy. & 3 &  0 &  1 & \underline{4}  \\
    & \textbf{5. Inaccurate parametric knowledge.} LLM-based verifiers make wrong verification due to the incorrect parametric knowledge stored in LLMs.  & 3 &  6 &  7 & 16 \\
    & \textbf{6. Insufficient or inaccurate externally collected knowledge} (evidence), involving three scenarios: (i) no external evidence, model makes wrong reasoning by itself; (ii) collected evidence is insufficient to cover all aspects mentioned by a long claim; (iii) collected evidence is inaccurate (e.g., evidence contain statement \textit{More than 430 species of mammal are found in the Amazon} when the correct number is 427). &  9 & 15 &  6 & \textbf{30}\\
    \midrule
    \multirow{6}{3cm}{III. LLM Reasoning} &
    \textbf{7. Incorrect reasoning}, e.g., the claim mentioned A while the model dismissed in the reasoning process, or the claim did not mention A while the model hallucinated A & 6 &  8 &  4 & 18 \\
    & \textbf{8. Strict reasoning} includes two situations: (i) strictly depending on the collect evidence to make decision leads to wrong verification, while if it combines commonsense and collected evidence to analyze, it can verify correctly; (2) strict reasoning based on parametric knowledge. E.g.\ regarding\textit{ Word-Cross Puzzle} and \textit{FUN Word-Cross Puzzle} are different. & 5 &  0 &  3 & 8 \\
    \midrule
    \multirow{3}{3cm}{IV. Debatable Opinion} &
    \textbf{9. Debatable Opinions} on controversial topics, e.g.\ actual origins are debated for claim \textit{Fortune cookies made their way to San Francisco in the late 1800s and early 1900s through Japanese immigrants.} & 7 &  8 &  0 & 15 \\
     \midrule
    \textbf{Total} &  & \textbf{66} & \textbf{45} & \textbf{24} & \textbf{135} \\ 
    \bottomrule
\end{tabular}
}
\caption{\textbf{Datasets Error distribution, grouped into nine fine-grained types under four major issues.}}
\label{tab:erroranalysis}
\end{table*}

\section{Error Analysis}
\label{sec-erroranalysis}

To identify weaknesses in our fact-checking system, we manually examine failed cases of three datasets: \felmwk, \factoolqa, and \bingcheck, analyzing whether the majority of failures is attributed to inadequate retrieved evidence or to flaws in the LLM verification process, despite the availability of reliable evidence.

We summarized errors into four major issues and nine error types.
Among the total number of 135 failed claims, there are 44 cases falling into challenges of (I) inaccurate identification of checkworthy claims and false gold labels in the original datasets, 50 claims are due to (II) inaccurate or insufficient knowledge applied to verification, either internally extracted from LLM parameters or externally collected from web pages. The rest 26 and 15 cases result from LLM reasoning ability and debatable opinions over some topics, respectively, as shown in ~\autoref{tab:erroranalysis}.

The major issue lies in collecting sufficient evidence, especially for long claims containing many aspects to verify. This can be approached by decomposing ``atomic claims'' from the original dataset into the real granularity of ``atomic'', each containing only 1-3 pieces of information.
The second problem focus on the quality of benchmarking datasets, particularly \felmwk that includes many ungrounded claims and labels~\citep{loki}, which may lead to ineffective comparisons between fact-checking systems.
Interestingly, beyond incorrect reasoning, overly-strict reasoning by exact matching between the claim and collected evidence can also lead to verification errors. For example, LLMs label a claim as false when the claim states \textit{FUN Word-Cross Puzzle} while evidence mentions \textit{Word-Cross Puzzle}. Additionally, some claims can be viewed as true from one perspective but false from another, as seen in debates over the origins of fortune cookies, where the truth of related claims is debatable.

Considering above, to further advance the field of fact-checking, we highlight the need for improved benchmarking datasets, a stronger focus on verifying fine-grained claims, and strategies to guide LLMs in performing verification under more flexible reasoning conditions, such as semantic alignment, rather than relying exclusively on exact matches.

%% file: sections/appendix.tex
\section{Prompts for Verification}
\label{appendix-verificationprompt}

\paragraph{Default prompt}
We use the following prompt to guide the language model in verifying the atomic claim, determining whether to provide a final judgment or issue an additional Google search query based on the current status. The prompt will output reason or explanation for the verification process.

\begin{lstlisting}
_FINAL_ANSWER_OR_NEXT_SEARCH_FORMAT = f"""Instructions:
1. You are provided with a STATEMENT and relevant KNOWLEDGE points.
2. Based on the KNOWLEDGE, assess the factual accuracy of the STATEMENT.
3. Before presenting your conclusion, think through the process step-by-step. Include a summary of the key points from the KNOWLEDGE as part of your reasoning.
4. If the KNOWLEDGE allows you to confidently make a decision, output the final answer as a JSON object in the following format:
   {{
     "final_answer": "{_Factual_LABEL}" or "{_Non_Factual_LABEL}"
   }}
5. If the KNOWLEDGE is insufficient to make a judgment, issue ONE Google Search query that could provide additional evidence. Output the search query in JSON format, as follows:
   {{
     "search_query": "Your Google search query here"
   }}
6. The query should aim to obtain new information not already present in the KNOWLEDGE, specifically helpful for verifying the STATEMENT's accuracy.

KNOWLEDGE:
{_KNOWLEDGE_PLACEHOLDER}

STATEMENT:
{_STATEMENT_PLACEHOLDER}
"""
\end{lstlisting}

\paragraph{No Reason prompt}
To improve efficiency, we opted for this setting to output only the label.

\begin{lstlisting}
_FINAL_ANSWER_OR_NEXT_SEARCH_FORMAT = f"""Instructions:
1. You are provided with a STATEMENT and relevant KNOWLEDGE points.
2. Based on the KNOWLEDGE, assess the factual accuracy of the STATEMENT.
3. Before presenting your conclusion, think through the process step-by-step. Include a summary of the key points from the KNOWLEDGE as part of your reasoning.
4. If the KNOWLEDGE allows you to confidently make a decision, output the final answer as a JSON object in the following format:
   {{
     "final_answer": "{_Factual_LABEL}" or "{_Non_Factual_LABEL}"
   }}
5. If the KNOWLEDGE is insufficient to make a judgment, issue ONE Google Search query that could provide additional evidence. Output the search query in JSON format, as follows:
   {{
     "search_query": "Your Google search query here"
   }}
6. The query should aim to obtain new information not already present in the KNOWLEDGE, specifically helpful for verifying the STATEMENT's accuracy.
7. Do not provide any additional information or reasoning in the output. Only output the JSON object.

KNOWLEDGE:
{_KNOWLEDGE_PLACEHOLDER}

STATEMENT:
{_STATEMENT_PLACEHOLDER}
"""
\end{lstlisting}

\paragraph{At Least One prompt}
At Least One prompt requires models to retrieve at least one evidence during the verification.
\begin{lstlisting}
_FINAL_ANSWER_OR_NEXT_SEARCH_FORMAT = f"""Instructions:
1. You are provided with a STATEMENT and relevant KNOWLEDGE points.
2. Based on the KNOWLEDGE, assess the factual accuracy of the STATEMENT.
3. Before presenting your conclusion, think through the process step-by-step. Include a summary of the key points from the KNOWLEDGE as part of your reasoning.
4. If the KNOWLEDGE allows you to confidently make a decision, output the final answer as a JSON object in the following format:
   {{
     "final_answer": "{_Factual_LABEL}" or "{_Non_Factual_LABEL}"
   }}
5. If the KNOWLEDGE is insufficient to make a judgment, issue ONE Google Search query that could provide additional evidence. Output the search query in JSON format, as follows:
   {{
     "search_query": "Your Google search query here"
   }}
6. The query should aim to obtain new information not already present in the KNOWLEDGE, specifically helpful for verifying the STATEMENT's accuracy.
7. If the KNOWLEDGE is empty, please issue ONE Google Search query immediately.

KNOWLEDGE:
{_KNOWLEDGE_PLACEHOLDER}

STATEMENT:
{_STATEMENT_PLACEHOLDER}
"""
\end{lstlisting}

\paragraph{At Least Two prompt}
At Least Two prompt is a more aggressive approach to retrieve minimum two evidence before verification.
\begin{lstlisting}
_FINAL_ANSWER_OR_NEXT_SEARCH_FORMAT = f"""Instructions:
1. You are provided with a STATEMENT and relevant KNOWLEDGE points.
2. Based on the KNOWLEDGE, assess the factual accuracy of the STATEMENT.
3. Before presenting your conclusion, think through the process step-by-step. Include a summary of the key points from the KNOWLEDGE as part of your reasoning.
4. If the KNOWLEDGE allows you to confidently make a decision, output the final answer as a JSON object in the following format:
   {{
     "final_answer": "{_Factual_LABEL}" or "{_Non_Factual_LABEL}"
   }}
5. If the KNOWLEDGE is insufficient to make a judgment, issue ONE Google Search query that could provide additional evidence. Output the search query in JSON format, as follows:
   {{
     "search_query": "Your Google search query here"
   }}
6. The query should aim to obtain new information not already present in the KNOWLEDGE, specifically helpful for verifying the STATEMENT's accuracy.
7. If the KNOWLEDGE is empty or there is only ONE evidence in the KNOWLEDGE, please issue ONE Google Search query immediately.

KNOWLEDGE:
{_KNOWLEDGE_PLACEHOLDER}

STATEMENT:
{_STATEMENT_PLACEHOLDER}
"""
\end{lstlisting}

\paragraph{Inclusive}
In this setting, we prompt models to be ``less strict, open-minded and avoid being over confident'' to encourage models to reflect on their confidence level of answers.

\begin{lstlisting}
_FINAL_ANSWER_OR_NEXT_SEARCH_FORMAT = f"""Instructions:
1. You are provided with a STATEMENT and relevant KNOWLEDGE points.
2. Based on the KNOWLEDGE, assess the factual accuracy of the STATEMENT.
3. Before presenting your conclusion, think through the process step-by-step. Include a summary of the key points from the KNOWLEDGE as part of your reasoning.
4. If the KNOWLEDGE allows you to confidently make a decision, output the final answer as a JSON object in the following format:
   {{
     "final_answer": "{_Factual_LABEL}" or "{_Non_Factual_LABEL}"
   }}
5. If the KNOWLEDGE is insufficient to make a judgment, issue ONE Google Search query that could provide additional evidence. Output the search query in JSON format, as follows:
   {{
     "search_query": "Your Google search query here"
   }}
6. The query should aim to obtain new information not already present in the KNOWLEDGE, specifically helpful for verifying the STATEMENT's accuracy.
7. Please be more open-minded and less strict in your evaluation. Avoid being overly confident, and consider the possibility of alternative interpretations or uncertainties in the evidence.

KNOWLEDGE:
{_KNOWLEDGE_PLACEHOLDER}

STATEMENT:
{_STATEMENT_PLACEHOLDER}
"""
\end{lstlisting}

\section{Prompt for Final Verification}
\label{appendix-finaldecision}

Upon reaching the maximum number of steps, we issue the following prompt to compel the language model to make a final judgment based on the accumulated information.

\begin{lstlisting}
_MUST_HAVE_FINAL_ANSWER_FORMAT = f"""Instructions:
1. You are provided with a STATEMENT and relevant KNOWLEDGE points.
2. Based on the KNOWLEDGE, assess the factual accuracy of the STATEMENT.
3. Before presenting your final answer, think step-by-step and show your reasoning. Include a summary of the key points from the KNOWLEDGE as part of your reasoning.
4. Your final answer should be either "{_Factual_LABEL}" or "{_Non_Factual_LABEL}".
5. Format your final answer as a JSON object in the following structure:
   {{
     "final_answer": "{_Factual_LABEL}" or "{_Non_Factual_LABEL}"
   }}
6. Do not include any other information or reasoning in the output. Only provide the JSON object.

KNOWLEDGE:
{_KNOWLEDGE_PLACEHOLDER}

STATEMENT:
{_STATEMENT_PLACEHOLDER}
"""
\end{lstlisting}

% \section{Dataset Statistics}
% \label{appendix-ds}

% \begin{table}[t]
% \centering
% \begin{tabular}{lccccc}
% \toprule
% \textbf{Dataset} & \textbf{\#SP} & \textbf{\#PS} & \textbf{\#NS} & \textbf{\#RF} & \textbf{Total} \\
% \midrule
% BingCheck & 2681 & 900 & 217 & 42 & 3840 \\
% \bottomrule
% \end{tabular}
% \caption{\textbf{Statistics of each labels in the original dataset.} SP indicates supported, PS indicates partially supported, NS indicates not supported, RF indicates refuted.}
% \label{tab-dataset-stats-origin}
% \end{table}

% \section{Search Query Repetition Analysis}
% \label{appendix-searchquery}

\section{Effect of Reasoning}
\label{appendix-reasoningeffect}

\begin{figure}[t]
     \centering
     \includegraphics[width=\columnwidth]{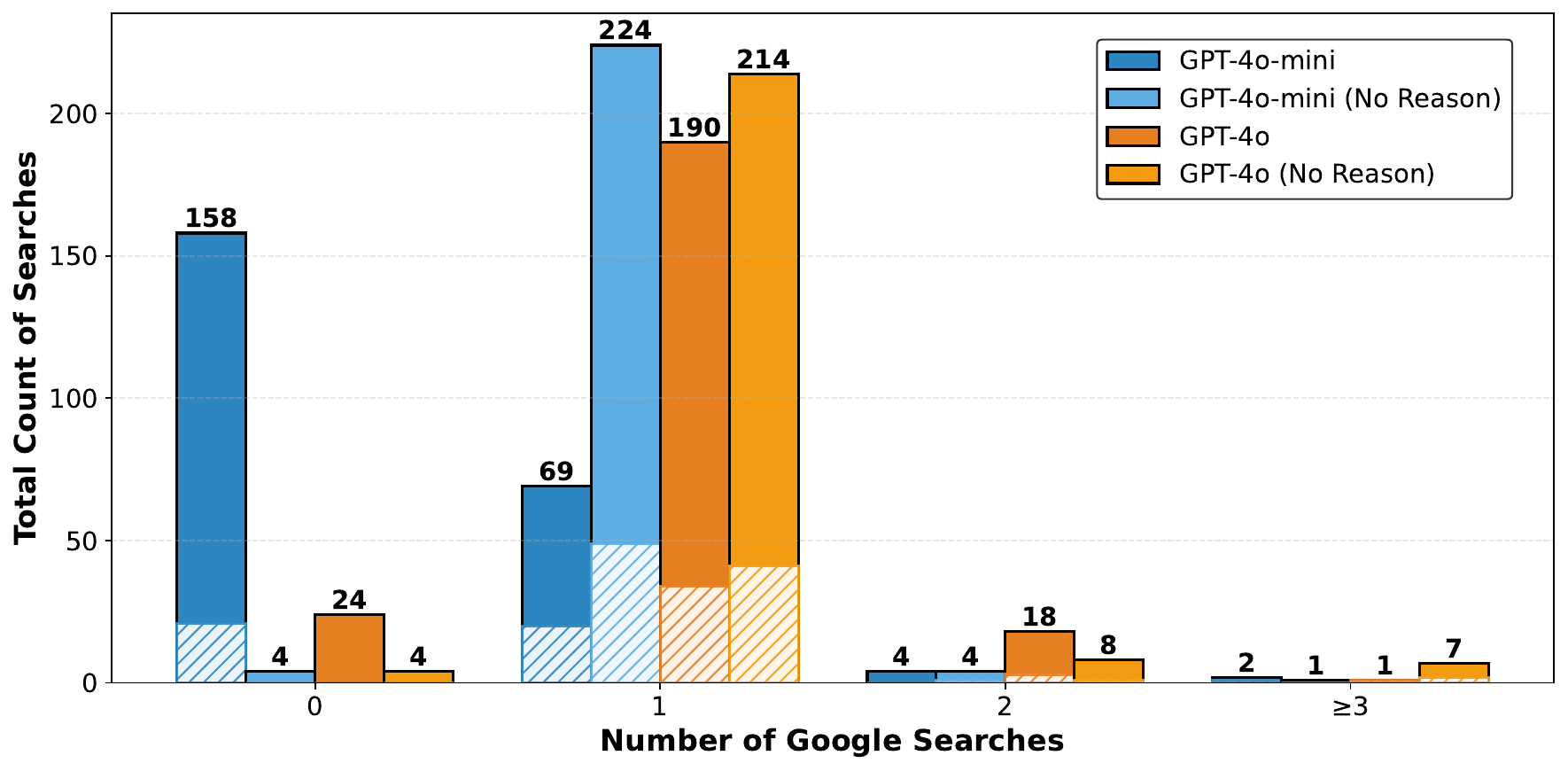}
        \caption{
        \textbf{The effect of reasoning on the number of searches using \fouro and \fouromini within \ourframework on \factoolqa.}
        }
        \label{fig-factoolqasearchnumber}
\end{figure}

\begin{figure}[t]
     \centering
     \includegraphics[width=\columnwidth]{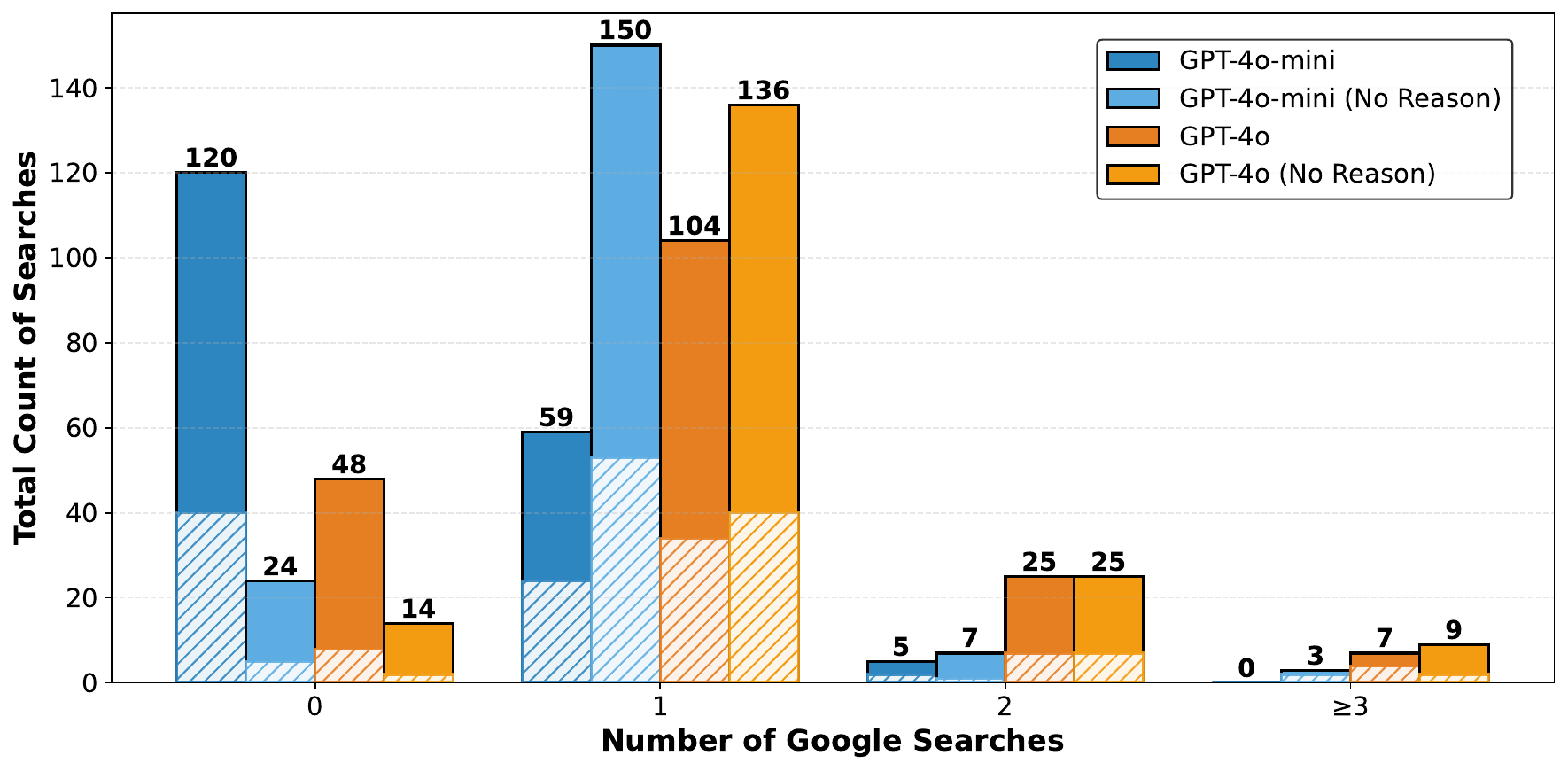}
        \caption{
        \textbf{The effect of reasoning on the number of searches using \fouro and \fouromini within \ourframework on \felmwk.}
        }
        \label{fig-felmwksearchnumber}
\end{figure}

We additionally include figures to illustrate the effect of reasoning on two other datasets: \factoolqa (\autoref{fig-factoolqasearchnumber}) and \felmwk (\autoref{fig-felmwksearchnumber}), supplementing \autoref{fig-bingchecksearchnumbers}. These figures demonstrate that both \fouro and \fouromini are influenced by explicitly stating their reasoning process, with \fouromini showing a consistent impact across all datasets, not just \bingcheck.
Furthermore, when comparing these datasets, we observe that the models appear most confident on \felmwk compared to the other two datasets. As a result, even in the absence of explicit reasoning, they do not perform any searches to verify the claims.

